\documentclass[preprint,12pt]{elsarticle}

\usepackage{graphicx}
\usepackage{bm}
\usepackage{color}
\usepackage{amssymb}
\usepackage{a4wide}
\usepackage{hyperref}

\begin{document}

\begin{frontmatter}

\title{Determination of the structure of the $X(3872)$ in $\bar p A$ collisions}

\author[lab1,lab2]{A.B. Larionov}
\ead{larionov@fias.uni-frankfurt.de}
\author[lab3]{M. Strikman}
\ead{strikman@phys.psu.edu}
\author[lab1,lab4]{M. Bleicher}
\ead{bleicher@th.physik.uni-frankfurt.de}

\address[lab1]{Frankfurt Institute for Advanced Studies (FIAS), 
               D-60438 Frankfurt am Main, Germany}
\address[lab2]{National Research Centre "Kurchatov Institute", 
               123182 Moscow, Russia}
\address[lab3]{Pennsylvania State University, University Park, PA 16802, USA}
\address[lab4]{Institut f\"ur Theoretische Physik, J.W. Goethe-Universit\"at,
             D-60438 Frankfurt am Main, Germany}

\date{\today}

\begin{abstract}
Currently, the structure of the $X(3872)$ meson is unknown. Different competing 
models of the  $c\bar c$ exotic state $X(3872)$ exist, including the possibilities
that this state is either a mesonic molecule with dominating $D^0 \bar D^{*0} +c.c.$ 
composition, a $c \bar c q \bar q$ tetraquark, or a $c \bar c$-gluon hybrid state. 
It is expected that the $X(3872)$ state is rather strongly coupled to the $\bar p p$ 
channel and, therefore, can be produced in $\bar p p$ and $\bar pA$ collisions at PANDA.
We propose to test the hypothetical molecular structure of $X(3872)$ by studying the 
$D$ or $\bar D^{*}$ stripping reactions on a nuclear residue.
\end{abstract}

\begin{keyword}

X(3872) \sep $\bar p A$ reactions \sep charmed meson production

\PACS 25.43.+t \sep 14.40.Rt \sep 14.40.Lb \sep 24.10.Ht

\end{keyword}

\end{frontmatter}

\section{Introduction}
\label{Intro}

The discovery of exotic $c \bar c$ mesons at B-factories and at the Tevatron
stimulated interest to explore the possible existence of tetraquark 
and molecular meson states. The famous $X(3872)$ state has been originally found by BELLE 
\cite{Choi:2003ue} as a peak in $\pi^+ \pi^- J/\psi$ invariant mass spectrum  from exclusive 
$B^{\pm} \to K^{\pm} \pi^+ \pi^- J/\psi$ decays. Nowadays the existence of the $X(3872)$ state
and its quantum numbers $J^{PC}=1^{++}$ are well established \cite{Agashe:2014kda}.
In particular, radiative decays $X(3872) \to J/\psi \gamma$, $X(3872) \to \psi^\prime(2S) \gamma$ 
\cite{Aubert:2008ae} point to the positive $C$-parity of the $X(3872)$. 
Probably the most intriguing feature is that the mass of the $X(3872)$ is within 1 MeV 
the sum of the $D^0$ and $D^{*0}$ meson masses. This prompted the popular conception of the $X(3872)$ 
being a $D \bar D^* + \bar D D^*$ molecule. 


To probe the molecular nature of the $X(3872)$ structure has been difficult.
So far, most theoretical calculations have been focused on the description
of radiative and isospin-violating decays of the $X(3872)$. For example, 
the $X(3872) \to J/\psi \gamma$ decay can be well understood within the $D \bar D^* + c.c.$ 
molecular hypothesis \cite{Aceti:2012cb}. On the other hand, the measured large branching fraction
$B(X(3872) \to \psi^\prime(2S) \gamma)/B(X(3872) \to J/\psi \gamma) = 3.4 \pm 1.4$
\cite{Aubert:2008ae} seems to disfavour the molecular structure and requires 
a significant pure $c \bar c$ admixture in the $X(3872)$ \cite{Swanson:2004pp}.
The theoretical predictions for the decay rates are, however, quite 
sensitive to the model details even within various approaches like charmonium or 
$D \bar D^*+c.c.$ molecular models.

In this letter we suggest to test the charm meson molecular hypothesis of the $X(3872)$
structure in $\bar p A$ collisions at PANDA. Assuming that the $X(3872)$ is coupled 
to the $p \bar p$ channel, we consider the stripping reaction of the $D$-meson on 
a nuclear target nucleon such that a $\bar D^{*}$ is produced and vice versa.
We show that the distribution of the produced charmed meson
in the light cone momentum fraction $\alpha$ with $z$-axis along $\bar p$ beam momentum, 
\begin{equation}
   \alpha=\frac{2(\omega_{D^*}(\mathbf{k}_{D^*})+k_{D^*}^z)}{E_{\bar p}+m_N+p_{\rm lab}}~, \label{alpha_Acm}
\end{equation}
will be sharply peaked at $\alpha \simeq 1$ at small transverse momenta
which allows to unambiguously identify the weakly coupled $D \bar D^* + c.c.$ molecule.
Here, $\mathbf{k}_{D^*}$ and $\omega_{D^*}(\mathbf{k}_{D^*})=(\mathbf{k}_{D^*}^2+m_{D^*}^2)^{1/2}$
are, respectively, the momentum and energy of the produced $\bar D^{*}$ meson 
in the target nucleus rest frame.
Similar studies of hadron-, lepton-, and nucleus-deuteron interactions at high energy
have been  proposed long ago to test the deuteron structure at short distances as 
in the spectator kinematics the $n$- or $p$-stripping cross sections are proportional 
to the square of the deuteron wave function. For the $X(3872)$ this idea is depicted 
in Fig.~\ref{fig:pDDbar_dia} (details follow below).

\section{$X(3872)$-proton cross section}
\label{elem}

For brevity, the bar, which can be seen over the $D^*$ or $D$, will be dropped in many cases below.
The charge conjugated states are implicitly included in the calculated cross sections.  
 
The most important ingredients of our calculations are the total $Xp$ cross section
and the momentum differential cross section $X p \to D^*(D) + {\rm anything}$. In the molecular picture,
the latter cross section is the $D (D^*)$-meson stripping cross section. 
\begin{figure}
\begin{center}
\begin{tabular}{ll}
\includegraphics[scale = 0.7]{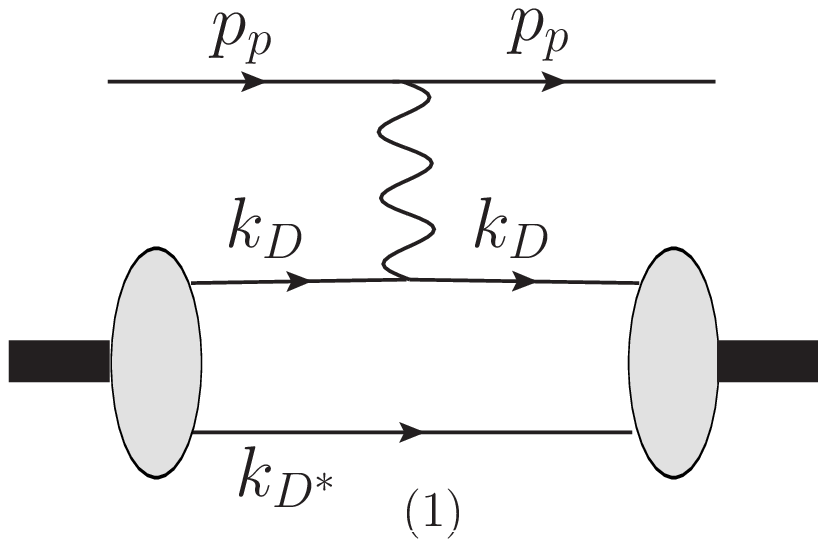} & 
\includegraphics[scale = 0.7]{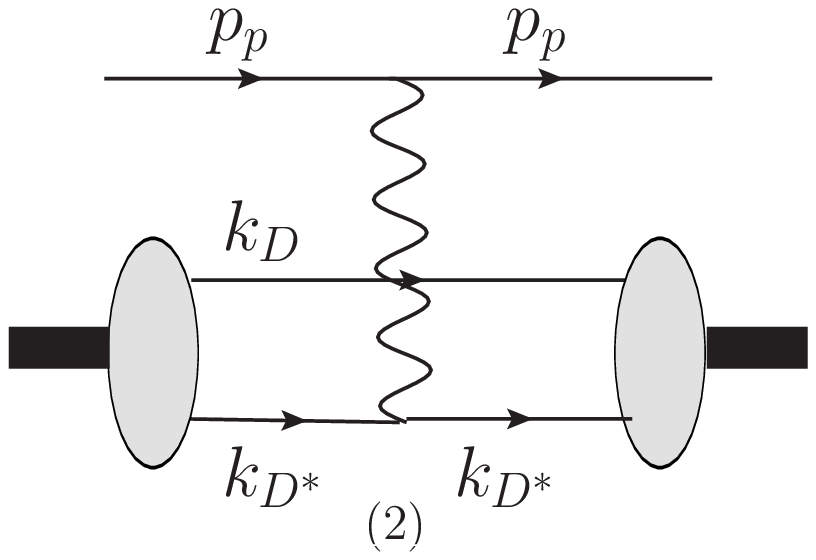} \\
\includegraphics[scale = 0.7]{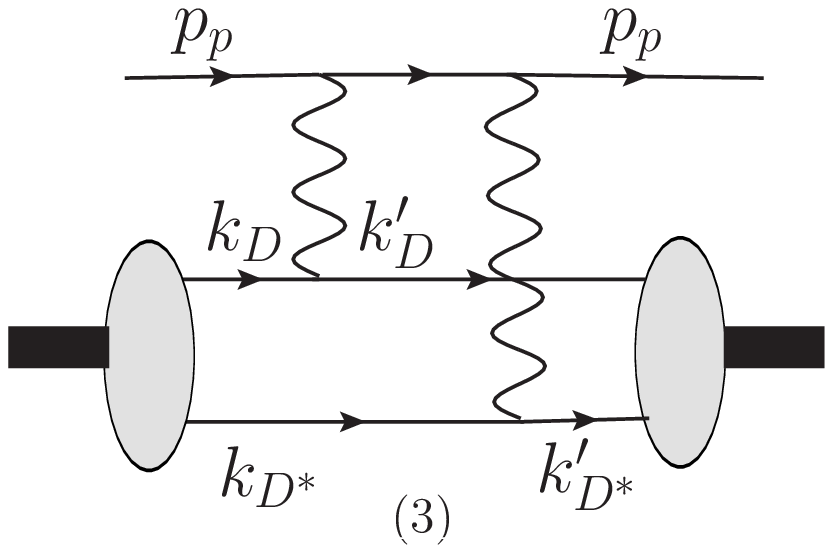} &
\includegraphics[scale = 0.7]{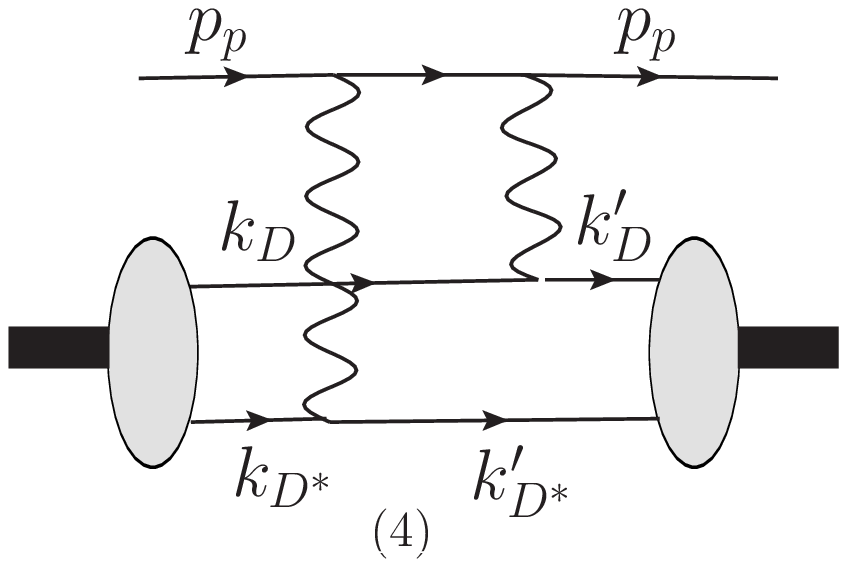}
\end{tabular}
\end{center}
\caption{\label{fig:pDDbar_dia} Processes contributing to the forward scattering
amplitude of a proton on the $DD^*$ molecule. 
Wavy lines denote the $pD$ and $pD^*$ elastic scattering amplitudes.
Straight lines are labeled with particle's four-momenta. The blobs represent the wave function
of the molecule.}
\end{figure} 
To calculate the total $Xp$ cross section within the Glauber theory, we start from the graphs shown 
in Fig.~\ref{fig:pDDbar_dia} which assume the $D D^*$ composition of $X(3872)$. 
It is convenient to perform calculations in the $DD^*$ 
molecule center-of-mass (c.m.) frame with proton momentum $\mathbf{p}_{\rm p}$ directed along $z$-axis.
The invariant forward scattering amplitudes of the first two processes are 
\begin{eqnarray}
   i M^{(1)}(0)  &=& \int d^3k \frac{m_X}{\omega_D} |\psi(\mathbf{k})|^2 i M_{pD}(0)~,    \label{M^(1)}  \\
   i M^{(2)}(0)  &=& \int d^3k \frac{m_X}{\omega_{D^*}} |\psi(\mathbf{k})|^2 i M_{pD^*}(0)~,  \label{M^(2)}
\end{eqnarray}
where $m_X=\omega_D+\omega_{D^*}$ is the mass of the molecule and $\omega_D$ ($\omega_{D^*}$) is the energy of $D$ ($D^*$)-meson.
(The different assumptions on the momentum dependence of meson energies discussed in the next section have practically
no effect on the $Xp$ cross section.) 
The molecule wave function in momentum space is defined as
\begin{equation}
   \psi(\mathbf{k}) = \int \frac{d^3 r}{(2\pi)^{3/2}} \mbox{e}^{-i\mathbf{k r}} 
                   \psi(\mathbf{r})~,   \label{psi(k)}
\end{equation}
where $\mathbf{k}$ is the $D^*$ momentum in the $D D^*$ c.m. frame, with
the normalization condition $\int d^3k |\psi(\mathbf{k})|^2=1$.

For the calculation of the third and forth
processes in Fig.~\ref{fig:pDDbar_dia} we apply the generalized eikonal 
approximation (GEA) \cite{Frankfurt:1996xx,Sargsian01} which assumes the nonrelativistic motion
of $D$ and $D^*$ inside the molecule. In this approximation, the propagator
of the intermediate proton depends only on the $z$-component of 
momentum transfer $\mathbf{q} \equiv \mathbf{k}_{D^*}-\mathbf{k}_{D^*}^\prime$,
while the $pD$ and $pD^*$ elastic scattering amplitudes depend only on the momenta of incoming particles
and on the transverse momentum transfer. Thus, we obtain
\begin{eqnarray}
   &&i M^{(3)}(0) =
   \int \frac{d^3 k d^3 q}{(2\pi)^3} \frac{i m_X}{2 \omega_D \omega_{D^*}} \psi^*(\mathbf{k}-\mathbf{q})
   \frac{i M_{pD^*}(\mathbf{q}_{t}) i M_{pD}(-\mathbf{q}_{t})}{2 p_{\rm p}(q^z+i\varepsilon)}
   \psi(\mathbf{k})~,     \label{M^(3)} \\
   &&i M^{(4)}(0) =
   \int \frac{d^3 k d^3 q}{(2\pi)^3} \frac{i m_X}{2 \omega_D \omega_{D^*}} \psi^*(\mathbf{k}-\mathbf{q})
   \frac{i M_{pD^*}(\mathbf{q}_{t}) i M_{pD}(-\mathbf{q}_{t})}{2 p_{\rm p}(-q^z+i\varepsilon)}
   \psi(\mathbf{k})~.    \label{M^(4)}
\end{eqnarray}
Therefore,
\begin{equation}
   iM^{(3)}(0)+ iM^{(4)}(0) =
   \int \frac{d^3 k d^2 q_{t}}{(2\pi)^2} \frac{m_X}{4 \omega_D \omega_{D^*} p_{\rm p}} \psi^*(\mathbf{k}-\mathbf{q}_{t}) i M_{pD^*}(\mathbf{q}_{t})
   i M_{pD}(-\mathbf{q}_{t}) \psi(\mathbf{k})~.    \label{M^(34)} 
\end{equation}
The optical theorem for the proton-molecule forward scattering amplitude is
\begin{equation}
    \mbox{Im}M(0) =  2p_{\rm p}m_X \sigma_{pX}^{\rm tot}~.   \label{OptTheoremElem1}
\end{equation}
Substituting $M(0)= M^{(1)}(0)+M^{(2)}(0)+M^{(3)}(0)+M^{(4)}(0)$
and using the parameterization of the strong interaction scattering amplitudes in the usual form as
\begin{equation}
   M_{pD^{(*)}}(\mathbf{q}_{t})= 2 i I_{pD^{(*)}}(\mathbf{k}_{D^{(*)}}) \sigma_{pD^{(*)}}^{\rm tot} \mbox{e}^{-B_{pD^{(*)}} q_{t}^2/2}~,   \label{M_pD_qt}
\end{equation}
with $I_{pD^{(*)}}(\mathbf{k}_{D^{(*)}}) =[(E_p \omega_{D^{(*)}}-p_{\rm p}k_{D^{(*)}}^z)^2-(m_p m_{D^{(*)}})^2]^{1/2}$ being the Moeller flux factor
we obtain the following expression for the proton-molecule total cross section:
\begin{eqnarray}
   &&\sigma_{pX}^{\rm tot} =  \int d^3k |\psi(\mathbf{k})|^2 
                                     [  {\cal I}_{pD}(-\mathbf{k}) \sigma_{pD}^{\rm tot}
                                      + {\cal I}_{pD^*}(\mathbf{k}) \sigma_{pD^*}^{\rm tot} ] 
    - \frac{1}{2} \int d^3 k \psi(\mathbf{k}) {\cal I}_{pD}(-\mathbf{k}) \sigma_{pD}^{\rm tot}
                                                 {\cal I}_{pD^*}(\mathbf{k}) \sigma_{pD^*}^{\rm tot}
                                                      \nonumber \\
   && \hspace{5cm} \times \int \frac{d^2 q_{t}}{(2\pi)^2} \psi^*(\mathbf{k}-\mathbf{q}_{t})
                           \mbox{e}^{-(B_{pD^*}+B_{pD})q_{t}^2/2}~,    \label{sigma_pX^tot}
\end{eqnarray}
where the normalized flux factors are defined as 
${\cal I}_{pD^{(*)}}(\mathbf{k}) \equiv I_{pD^{(*)}}(\mathbf{k})/p_{\rm p}\omega_{D^{(*)}}$.
In the small binding energy limit the molecule wave function decreases rapidly with increasing 
momentum $k$ and becomes negligibly small at $k \ll B_{pD}^{-1/2}$. In this case one can set 
$B_{pD} = B_{pD^*} = 0$ and perform the Taylor expansion of the flux factors in $k^z$ 
in Eq.(\ref{sigma_pX^tot}). Then, for the S-state molecule with accuracy up to the linear terms 
in $k^z/m_D$ and assuming that $m_D \simeq m_D^*$, $\sigma_{pD^*}^{\rm tot} \simeq \sigma_{pD}^{\rm tot}$ 
we obtain the formula
\begin{equation}
     \sigma_{pX}^{\rm tot} = \sigma_{pD^*}^{\rm tot} + \sigma_{pD}^{\rm tot}
      - \frac{\sigma_{pD^*}^{\rm tot}\sigma_{pD}^{\rm tot}}{4\pi}
         \langle r^{-2} \rangle_{DD^*}~,                  \label{sigma_pX^tot_deuteron}
\end{equation}
in line with previous calculations of the proton-deuteron total cross section \cite{Franco:1965wi}.

We choose the wave function of a $DD^*$ molecule as the asymptotic solution
of the Schroedinger equation at large distances:
\begin{equation}
   \psi(\mathbf{r})=\sqrt{\frac{\kappa}{2\pi}} \frac{\mbox{e}^{-\kappa r}}{r}~,  \label{psi_r}
\end{equation}
where the range parameter $\kappa=\sqrt{2 \mu E_b}$ depends on the reduced mass $\mu=m_D m_{D^*}/(m_D+m_{D^*})$
and on the binding energy $E_B$ of the molecule. The corresponding momentum space wave function is
\begin{equation}
   \psi(\mathbf{k})=\frac{\kappa^{1/2}/\pi}{\kappa^2+\mathbf{k}^2}~.  \label{psi_k}
\end{equation}

Let us now discuss the input parameters of our model.
Since there is no experimental information on $Dp$ and $D^*p$ interactions, we rely on simple estimates 
in the high-energy limit. 
For small-size $q\bar q$ configurations the color dipole model predicts the scaling of the 
total meson-nucleon cross section with the average
square of the transverse distance between quark and antiquark in the meson,  which is proportional
to the square of the Bohr radius $r_B=3/4\mu\alpha_s$. Here, $\mu = m_q m_{\bar q}/(m_q+m_{\bar q})$ is 
the reduced mass with $m_q$ and $m_{\bar q}$ being the constituent quark and antiquark masses.
The Bohr radii of pion, kaon, $D$-meson and $J/\psi$ are ordered as $r_{B\pi} > r_{BK} > r_{BD} > r_{BJ/\psi}$. 
Hence, we expect that the total meson-nucleon cross sections follow the same order. 
At a beam momentum of 3.5 GeV/c (1/2 of the momentum of $X(3872)$ formed in the $\bar p p \to X$ 
process on the proton at rest) the total $\pi^+ p$ and $K^+ p$ cross sections are about 28 mb and 17 mb, 
respectively \cite{Agashe:2014kda}.
The $J/\psi p$ cross section is expected to be much smaller, $3.5-6$ mb (c.f. \cite{Larionov:2013axa} 
and refs. therein.).
We assume the total $Dp$ cross section $\sigma_{pD}^{\rm tot}=14$ mb, i.e. slightly below 
the $K^+p$ total cross section.  This choice is in reasonable agreement with effective field theory 
calculations \cite{Tolos:2013kva}.

It is well known that at incident energies of a few GeV, the amplitude of meson (nucleon) - nucleon
elastic scattering is (to a good approximation) proportional to the product of the electric form factors
of the colliding hadrons (see e.g.\cite{Gerland:2005ca} and refs. therein).
Thus, in the exponential approximation for the $t$-dependence of the form factors, the slope parameters
$B_{pM}$ of the transverse momentum dependence 
of the meson-proton cross section at small $t$ should be proportional to $\langle r^2 \rangle_p+\langle r^2 \rangle_M$,  
where $\langle r^2 \rangle_p$ 
and $\langle r^2 \rangle_M$ are the mean-squared charge radii of the proton and meson, respectively. 
Since $\langle r^2 \rangle_M \propto r_{BM}^2$, the slope parameters should be also ordered as the Bohr radii. 
Empirical values at $p_{\rm lab}=3.65$ GeV/c are $B_{p\pi^+}=6.75 \pm 0.12 $ GeV$^{-2}$ and  
$B_{pK^+}=4.12 \pm 0.12 $ GeV$^{-2}$  as fitted at $0.05 \leq -t \leq 0.44$ GeV$^2$ \cite{Ambats:1973tp}. 
On the other hand, $B_{pJ/\psi}=3$ GeV$^{-2}$ at the comparable beam momenta \cite{Gerland:2005ca}. 
We will assume the value $B_{pD}= 4$ GeV$^{-2}$, since the Bohr radii of kaon and $D$-meson differ
by $\sim 30$\% only.  For the $pD^*$ interaction we assume
for simplicity $\sigma_{pD^*}^{\rm tot} = \sigma_{pD}^{\rm tot}$ and $B_{pD^*} = B_{pD}$.

Our educated guess on the $D$- and $D^*$-meson-nucleon cross sections and slope parameters should of course  
be checked experimentally.
The empirical information on $\sigma_{pD}^{\rm tot}$ can be obtained by measuring the $A$-dependence of the transparency
ratio of $D$-meson production in $\bar pA$ reactions at beam momenta beyond the charmonium resonance peaks,
where the background $\bar p p \to \bar D D$ channel dominates. The slope parameter $B_{pD}$ can be addressed by
measuring the transverse momentum spread of $D$-meson production in $\bar pA$ reactions.

We will further assume that the $X(3872)$ wave function contains $86\%$ of $D^0 \bar D^{*0} +c.c.$ 
and $12\%$ of the $D^+ D^{*-} +c.c.$ component as predicted by the local hidden gauge approach
\cite{Aceti:2012cb}. The binding energy of $D^0 \bar D^{*0}$ 
is likely less than 1 MeV \cite{Brambilla:2010cs} and can not be determined from existing data 
\cite{Agashe:2014kda} accurately enough. We set $E_b^{D^0 \bar D^{*0}}=0.5$ MeV 
and $E_b^{D^+ D^{*-}}=8$ MeV in numerical calculations. This corresponds to the range parameters
$\kappa_{D^0 \bar D^{*0}}=0.16$ fm$^{-1}$ and $\kappa_{D^+ D^{*-}}=0.64$ fm$^{-1}$.
With these parameters the total $pX$ cross section (\ref{sigma_pX^tot}) is
$\sigma_{pX}^{\rm tot}=26$ and $23$ mb for $D^0 \bar D^{*0}$ and $D^+ D^{*-}$ components,
respectively, at the molecule momentum of 7 GeV/c in the proton rest frame. 

\section{$D(\bar D^*)$ stripping cross section}

\begin{figure}
\includegraphics[scale = 0.7]{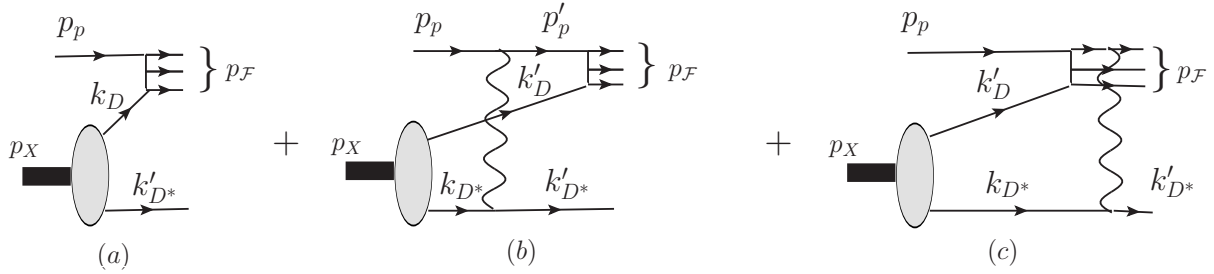}
\caption{\label{fig:pDDbar_to_F} The amplitude for the process $X(3872)+p \to D^* + {\cal F}$ where ${\cal F} \equiv \{{\cal F}_1,\ldots,{\cal F}_n\}$
is an arbitrary final state in the $pD$ interaction. See also caption to Fig.~\ref{fig:pDDbar_dia}.}
\end{figure}

In high energy hadron-deuteron reactions, the main contribution to the fast backward
nucleon production (in the deuteron rest frame or equivalently  - fast forward in the deuteron projectile case)
is given by the inelastic interaction of the hadron with second  nucleon of the deuteron \cite{Frankfurt:1976rk}.
For large nucleon momenta the spectrum is modified as compared to the impulse approximation (IA) due to 
the Glauber screening and antiscreening corrections \cite{Frankfurt:1979sv} since the hadron may interact 
with both nucleons. In a similar way, in calculations of the cross section $X+p \to D^* + {\rm anything}$, 
we take into account the IA diagram (Fig.~\ref{fig:pDDbar_to_F}a) and the single-rescattering diagrams of the incoming proton
(Fig.~\ref{fig:pDDbar_to_F}b) and of the outgoing proton or of the most energetic forward going baryon emerging 
from the inelastic $pD$ interaction (Fig.~\ref{fig:pDDbar_to_F}c).
The expressions for the invariant matrix elements for the processes (a) and (b) in Fig.~\ref{fig:pDDbar_to_F}
are straightforward to obtain in the c.m. frame of the molecule state $X$:
\begin{eqnarray}
   &&M^{(a)} = \sqrt{\frac{2m_X\omega_{D^*}}{\omega_D}} (2\pi)^{3/2} 
                M_{{\cal F};pD} \psi(\mathbf{k})~,  \label{M^(a)} \\
   &&M^{(b)} =  \frac{i m_X^{1/2}}{2p_p \sqrt{2\omega_D\omega_{D^*}}}
                \int d^3 r \psi(\mathbf{r}) \Theta(-z) \int \frac{d^2 q_{t}}{(2\pi)^2}
                \mbox{e}^{-i(\mathbf{k}+\mathbf{q}_{t})\mathbf{r}}       
                M_{{\cal F};p^\prime D^\prime} M_{pD^*}(\mathbf{q}_{t})~,  \label{M^(b)}
\end{eqnarray}
where $\mathbf{k} \equiv \mathbf{k}_{D^*}^\prime$. In the case of $M^{(b)}$ we applied the GEA
by expressing the propagator of the intermediate proton in the eikonal form and using the 
coordinate representation with $\mathbf{r}=\mathbf{r}_{D^*}-\mathbf{r}_{D}$.
The explicit form of the amplitude $M^{(c)}$ can be written only for specific outgoing states ${\cal F}$.
However, for the diffractive states including the leading proton, the expression 
for $M^{(c)}$ can be obtained from the expression for $M^{(b)}$ by replacing $\Theta(-z) \to \Theta(z)$,
which reflects the change of the time order of the $pD^*$ and $pD$ interactions.
Thus, for the diffractive outgoing state ${\cal F}$ the expression for $M^{(b)}+M^{(c)}$ is given by Eq. (\ref{M^(b)}) 
with replacement $\Theta(-z) \to 1$ (neglecting small differences
in momenta of incoming and outgoing proton in elementary amplitudes). 
We assume that the same replacement can be done for {\it any} 
final state ${\cal F}$. By summing over all states ${\cal F}$ we then obtain the momentum differential 
$D^*$ production (i.e. $D$-stripping) cross section in the molecule rest frame:
\begin{eqnarray}
   d \sigma_{p X \to D^*} &=& \frac{d^3 k_{D^*}^\prime}{(2\pi)^3 2 \omega_{D^*} 4p_pm_X}
   \sum_{\rm spins~and~sorts~of~{\cal F}}\int |M^{(a)}+M^{(b)}+M^{(c)}|^2 (2\pi)^4   \nonumber \\
 && \times \delta^{(4)}(p_{\cal F}+k_{D^*}^\prime-p_p-p_X) 
    \frac{d^3 p_{{\cal F}_1}}{(2\pi)^3 2E_{{\cal F}_1}} \cdots \frac{d^3 p_{{\cal F}_n}}{(2\pi)^3 2E_{{\cal F}_n}}~,  \label{dsigma_D^*_prelim}
\end{eqnarray}
where $p_X$ is the four momentum of the molecule ($p_X^2=m_X^2$). With a help of the unitarity 
relation for the elementary amplitudes \cite{BLP} the sum over spin states and sorts of ${\cal F}$ 
and the integration over phase space volume can be reduced to the products of the imaginary parts of 
elastic scattering amplitudes. This leads to the following expression for the momentum differential 
cross section in the molecule rest frame:
\begin{eqnarray}
   \frac{d^3 \sigma_{p X \to D^*}}{d^3 k} &=&
      \sigma_{pD}^{\rm tot} {\cal I}_{pD}(-\mathbf{k}) |\psi(\mathbf{k})|^2 \kappa  \label{dsigma_D^*}~,\\
   \kappa &=& 1-\sigma_{pD^*}^{\rm tot} {\cal I}_{pD^*}(\mathbf{k})  
            \int \frac{d^2 q_{t}}{(2\pi)^2} 
             \frac{\psi^*(\mathbf{k}+\mathbf{q}_{t})}{\psi^*(\mathbf{k})}
             \mbox{e}^{-(B_{pD}+B_{pD^*})\mathbf{q}_{t}^2/2}  \nonumber \\
    && +\frac{(\sigma_{pD^*}^{\rm tot} {\cal I}_{pD^*}(\mathbf{k}))^2}{4} 
        \int \frac{d^2q_{t} d^2q_{t}^\prime}{(2\pi)^4}
        \frac{\psi(\mathbf{k}+\mathbf{q}_{t}) \psi^*(\mathbf{k}+\mathbf{q}_{t}^\prime)}%
             {|\psi(\mathbf{k})|^2} \nonumber \\
    && \hspace{4cm} \times \mbox{e}^{-[B_{pD^*}(\mathbf{q}_{t}^2+\mathbf{q}_{t}^{\prime 2})
                  +B_{pD}(\mathbf{q}_{t}^\prime-\mathbf{q}_{t})^2]/2}~.  \label{kappa}     
\end{eqnarray} 
The first term in the r.h.s. of Eq.(\ref{kappa}) is the pure IA contribution. The second and third
terms are, respectively, the screening and antiscreening corrections (see Eqs. (8a) and (8b) in 
\cite{Frankfurt:1979sv}). 
The $D^*$ meson is assumed to be on its vacuum mass shell, 
$\omega_{D^*}(\mathbf{k})=\sqrt{m_{D^*}^2+\mathbf{k}^2}$,
while the energy of the $D$ meson is calculated from energy conservation, 
$\omega_D(-\mathbf{k})=m_X-\omega_{D^*}(\mathbf{k})$. 
(The condition $\omega_D > 0$ constrains the maximum momentum of the emitted $D^*$, 
$k < 3.3$ GeV/c. Above this value our model looses its applicability.)
In the case of the $D$-meson production one has to exchange ${\cal I}_{pD} \leftrightarrow {\cal I}_{pD^*}$,
$\sigma_{pD}^{\rm tot} \leftrightarrow \sigma_{pD^*}^{\rm tot}$ and $B_{pD} \leftrightarrow B_{pD^*}$ 
in Eqs.(\ref{dsigma_D^*}),(\ref{kappa}). In this case the on-shell condition is applied to the $D$-meson, 
while the $D^*$ energy is determined by energy conservation. 

It is convenient to express the differential invariant  $D^*$ production cross section 
(\ref{dsigma_D^*}) in terms of the relative fraction $\alpha$ of the light cone momentum 
of the $D D^*$ molecule carried by the $D^*$:
\begin{equation}
   \omega_{D^*}\frac{d^3 \sigma_{p X \to D^*}}{d^3 k}
   = \alpha \frac{d^3 \sigma_{p X \to D^*}}{d\alpha d^2 k_t}
   \equiv G_X^{p \to D^*}(\alpha,\mathbf{k}_t)~,    \label{dsigma_D^*_cov}
\end{equation}
where $\alpha=2(\omega_{D^*}(\mathbf{k})-k^z)/m_X$.
\begin{figure}
\includegraphics[scale = 0.7]{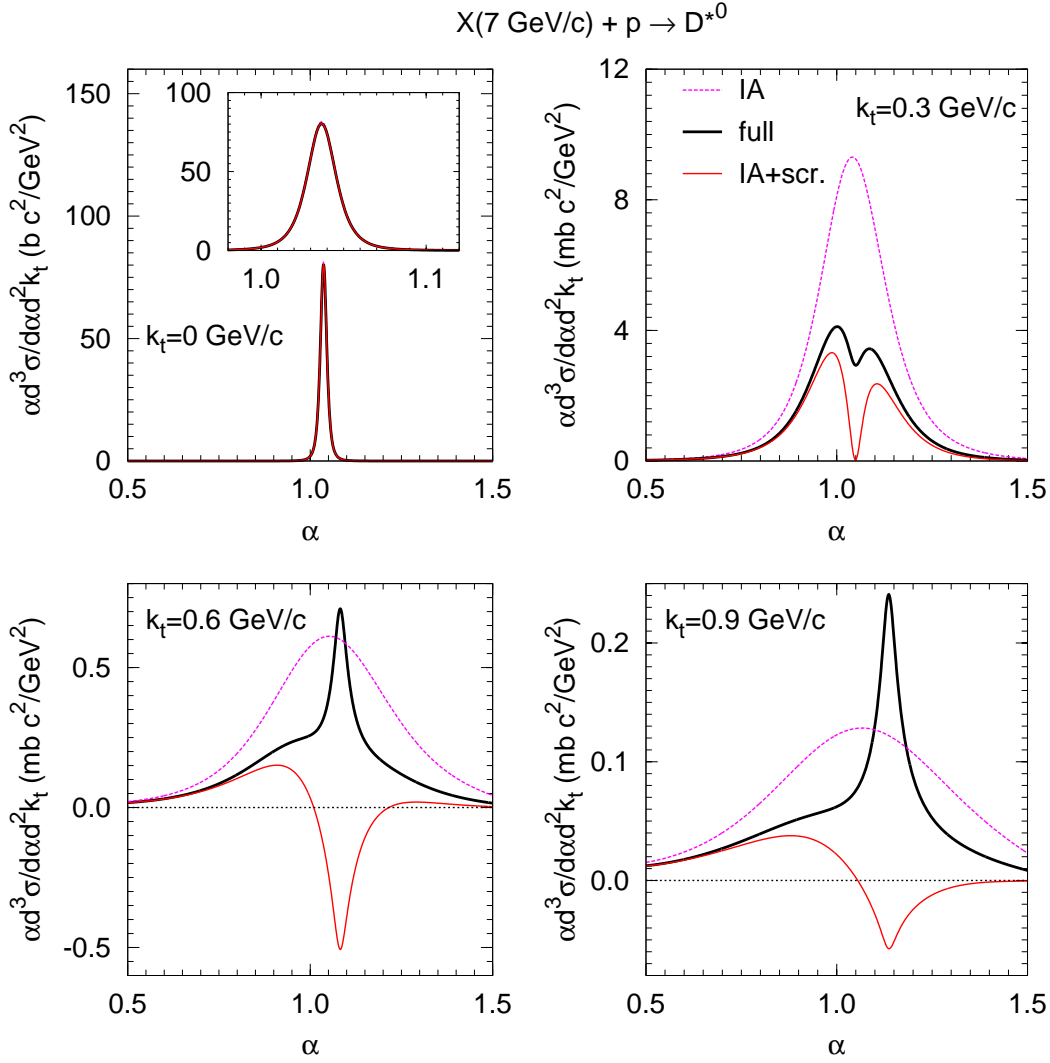}
\caption{\label{fig:Xp_to_Ds0} The invariant differential
cross section of $D^{*0}$ production in $X(3872)p$ collisions at $p_{\rm lab}=7$ GeV/c.
Thick solid line -- full calculation according to Eqs.(\ref{dsigma_D^*})-(\ref{dsigma_D^*_cov}).
Thin solid line -- the calculation taking into account only IA and screening term of Eq.(\ref{kappa}). 
Dashed line -- the calculation with $\kappa=1$ in Eq.(\ref{dsigma_D^*}), i.e. only with the IA term.
The inset at $k_t=0$ shows the behaviour of the differential cross section for a 
smaller range of $\alpha$.}
\end{figure}
\begin{figure}
\includegraphics[scale = 0.7]{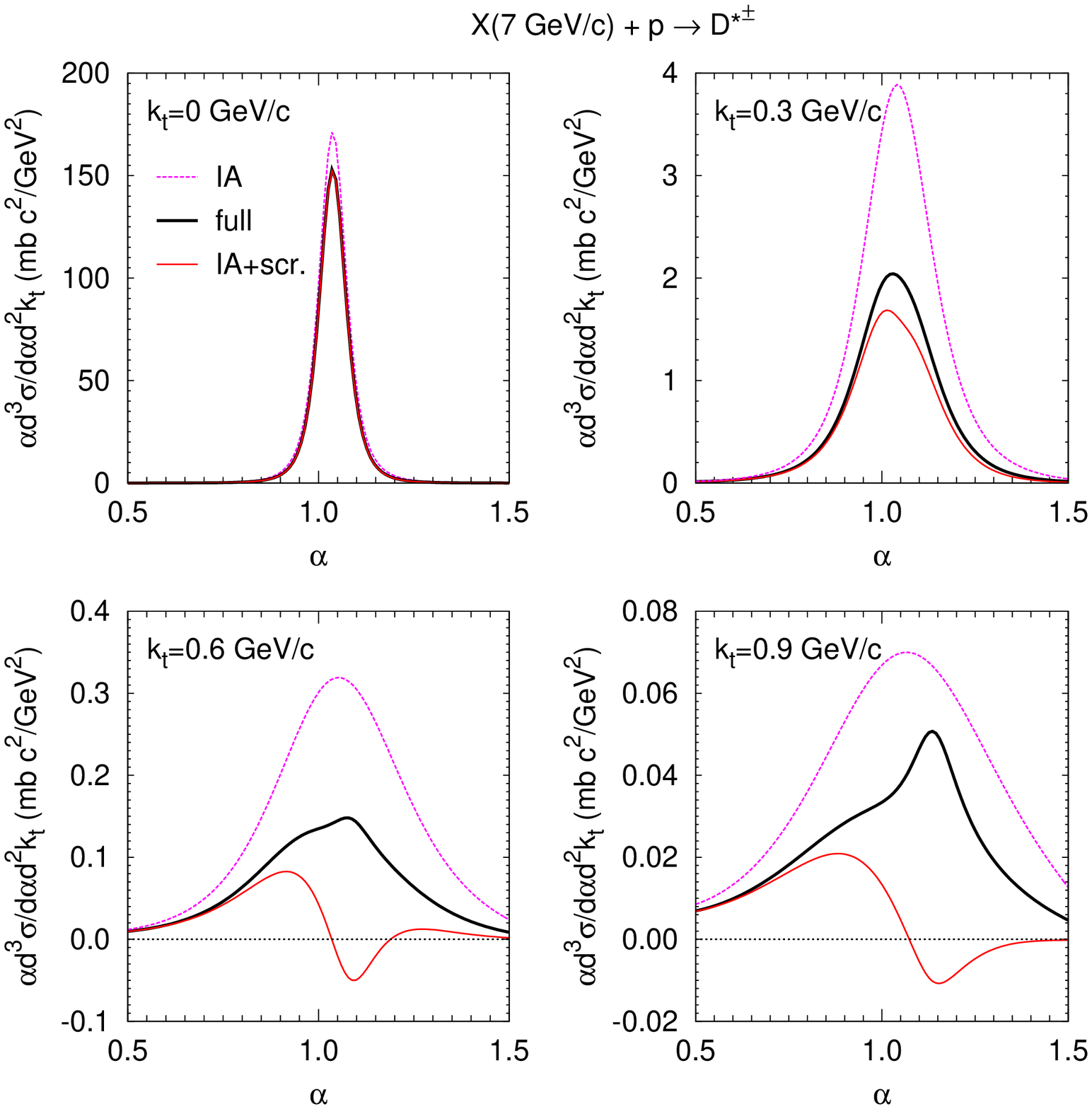}
\caption{\label{fig:Xp_to_Dsm} Same as Fig.~\ref{fig:Xp_to_Ds0} but for $D^{*\pm}$ 
production.}
\end{figure}
Figures \ref{fig:Xp_to_Ds0} and \ref{fig:Xp_to_Dsm} show the differential cross section
of $D^{*0}$ and $D^{*\pm}$ production from $X(3872)$ collisions at 7 GeV/c
with proton at rest as a function of $\alpha$ for several values of transverse momentum $k_t$.
At $k_t=0$, the cross section has a sharp maximum at $\alpha \simeq 2m_{D^*}/m_X \simeq 1.04$  
and  is almost unaffected by the screening and antiscreening corrections.
With increasing $k_t$, the width of $\alpha$-distribution
increases while the screening and antiscreening corrections to the IA term become important.
This is expected since the large-$k_t$ component of the molecule wave function corresponds
to small transverse separation between $D$ and $D^*$.
The corrections become large for $\alpha \simeq 1$ and large transverse momenta  as can be directly 
seen from the structure of the integrands in Eq.(\ref{kappa}). Indeed, $\alpha \simeq 1$ corresponds
to $k^z \simeq 0$ in the molecule rest frame. Then at finite transverse momentum transfer $q_{t}$
the ratio $\psi^*(\mathbf{k}_t+\mathbf{q}_{t})/\psi^*(\mathbf{k}_t)$ is less than unity at $k_t=0$ 
and asymptotically tends to unity with growing $k_t$. Due to the extremely narrow wave function of the
$D^0 D^{*0}$ molecule in momentum space, the screening and antiscreening corrections are sharply
peaked at $\alpha \simeq 1.1$ and develop structures in the $\alpha$-dependence of the cross section
at large transverse momenta. In the case of $\bar p A$ reactions these structures are slightly smeared out
due to the nucleon Fermi motion (see Fig.~\ref{fig:pbarA_to_Ds} below).

\section{$D^*$ and $D$ production off nucleus}
\label{pbarA}

\begin{figure}
\includegraphics[scale = 0.5]{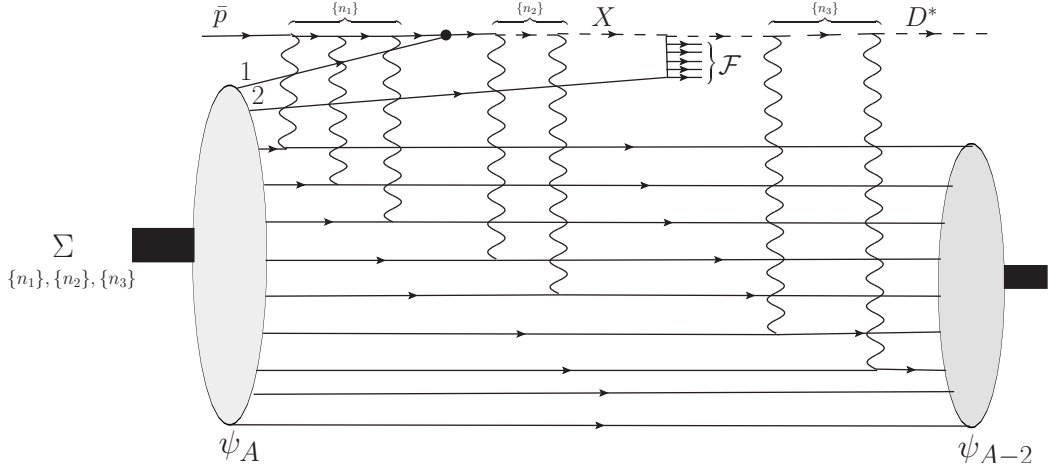}
\caption{\label{fig:pbarDs} The amplitude of the process $\bar p A \to D^* {\cal F} (A-2)^*$.
The wave functions of the initial and final nuclei are denoted as $\psi_A$ and $\psi_{A-2}$,
respectively. The mass numbers are shown as subscripts.
Wavy lines represent elastic scattering amplitudes on nucleons.
"${\cal F}$" stands for the arbitrary final state particles in the semi-inclusive process $X 2 \to D^* {\cal F}$.
The summation is performed over all possible sets of nucleon scatterers
$\{n_1\}$,$\{n_2\}$ and $\{n_3\}$ for the $\bar p$, $X$ and $D^*$, respectively.}
\end{figure}

In antiproton-nucleus interactions, we focus on the $D^*$ (or $D$) meson production 
in the two-step process $\bar p p \to X,~X N \to D^*(D) + {\rm anything}$. 
Similar to the case of $Xp$ interactions, we apply the Glauber theory to 
calculate the differential cross sections of the $D^*$ production in antiproton-nucleus 
interactions.
We start from the multiple scattering diagram  shown in Fig.~\ref{fig:pbarDs} which
can be evaluated within the GEA. We will assume that the nucleus can be described within 
the independent particle model disregarding the c.m. motion corrections (c.f. \cite{Larionov:2013nga}).
The incoming antiproton, intermediate molecular state $X$ and outgoing $D^*$-meson are allowed
to rescatter on nucleons elastically an arbitrary number of times.
The $D^*$ production cross section is proportional to the product of the sum of the amplitudes 
of Fig.~\ref{fig:pbarDs} and their conjugated.
The $X$ state is formed on a proton 1, while 
the $D^*$ is produced in the collision of $X$ with a nucleon 2. The nucleons 1 and 2 are fixed
in the direct and conjugated amplitudes while the sets of other nucleon scatterers are arbitrary.
The leading order contribution is given by the product term without elastic rescatterings.
Nuclear absorption corrections are accounted for by summing all possible product terms
with non-overlapping sets of nucleon scatterers. This gives the following expression for the momentum 
differential cross section of $D^*$ production on the nucleus:
\begin{eqnarray}
         \alpha \frac{d^3\sigma_{\bar p A \to D^*}}{d\alpha d^2k_t} &=&
 v_{\bar p}^{-1} \int d^3 r_1 {\cal P}_{\bar p,{\rm surv}}(\mathbf{b}_1,-\infty,z_1)
                 \int d^2 p_{1t} \frac{d^2\Gamma_{\bar p}^{1 \to X}}{d^2p_{1t}}
                 G_X^{p \to D^*}(\alpha,\mathbf{k}_t-\frac{\alpha}{2}\mathbf{p}_{1t}) \nonumber \\  
   && \times \int\limits_{z_1}^{\infty}dz_2 {\cal P}_{X,{\rm surv}}(\mathbf{b}_1,z_1,z_2)
             \rho(\mathbf{b}_1,z_2) {\cal P}_{D^*,{\rm surv}}(\mathbf{b}_1,z_2,\infty)~,
                                                                  \label{dsigma_barpA_to_DsX}
\end{eqnarray}
where
\begin{equation}
   \frac{d^2\Gamma_{\bar p}^{1 \to X}}{d^2p_{1t}} 
         = \frac{\overline{|M_{X;\bar p 1}|^2}\,v_{\bar p}}{(2\pi)^2 4 p_{\rm lab}^2 E_1}
            n_p(\mathbf{r}_1;\mathbf{p}_{1t},\Delta_{m_X}^0)     \label{Gamma_barp_to_X}
\end{equation}
is the in-medium width of $\bar p$ with respect to production of $X$ with transverse
momentum $\mathbf{p}_{1t}$; $v_{\bar p}=p_{\rm lab}/E_{\bar p}$ is the antiproton velocity;
$n_p(\mathbf{r}_1;\mathbf{p}_{1t},\Delta_{m_X}^0)$ is the proton occupation number;
$E_1=m_N-B$ with $B = 8.6$ MeV being the nucleon binding energy in $^{40}$Ar nucleus. 
The longitudinal momentum $\Delta_{m_X}^0$ of the proton 1 is obtained from the condition 
of on-shell production of the state $X$ in the process $\bar p~1 \to X$:
\begin{equation}
   \Delta_{m_X}^0 = \frac{m_N^2+E_1^2+2E_{\bar p}E_1-m_X^2}{2p_{\rm lab}}~.    \label{DeltaX^0}
\end{equation}
The nucleon occupation numbers are taken as the depleted Fermi distributions supplemented by
high-momentum tail due to short-range quasideuteron correlations (SRCs) 
\cite{Frankfurt:1981mk,Frankfurt:2008zv,Hen:2014nza}:
\begin{equation}
   n_q(\mathbf{r};\mathbf{p}) = (1-P_{2,q}) \Theta(p_{F,q}-p) 
            + \frac{\pi^2 P_{2,q} \rho_q  |\psi_d(p)|^2 \Theta(p-p_{F,q})}%
{\int\limits_{p_{F,q}}^\infty dp^\prime p^{\prime 2} |\psi_d(p^\prime)|^2}~,~~~~~~q=p,n \label{nWithTail}
\end{equation}
where $p_{F,q}(\mathbf{r})=[3\pi^2\rho_q(\mathbf{r})]^{1/3}$ are the nucleon Fermi momenta,
$P_{2,p} = 0.25$ and $P_{2,n} = P_{2,p} Z/N$ are the proton and neutron fractions above the Fermi surface, 
$\rho_q(\mathbf{r})$ are the nucleon densities, and $\psi_d(p)$ is the deuteron wave function.
In Eq.(\ref{dsigma_barpA_to_DsX}), the nuclear absorption is given  by the survival probabilities 
of the antiproton, the molecule, and the $D^*$: 
\begin{eqnarray}
   {\cal P}_{\bar p,{\rm surv}}(\mathbf{b}_1,-\infty,z_1) &=&
      \exp\left\{-\sigma_{p \bar p}^{\rm tot}
                  \int\limits_{-\infty}^{z_1} dz \rho(\mathbf{b}_1,z)\right\}~,
                                                                                 \label{Psurvpbar} \\
   {\cal P}_{X,{\rm surv}}(\mathbf{b}_1,z_1,z_2) &=&
      \exp\left\{-\sigma_{pX}^{\rm tot}
                  \int\limits_{z_1}^{z_2} dz \rho(\mathbf{b}_1,z)\right\}~,
                                                                                 \label{PsurvX} \\
   {\cal P}_{D^*,{\rm surv}}(\mathbf{b}_1,z_2,\infty)  &=&
      \exp\left\{-\sigma_{pD^*}^{\rm tot}
                  \int\limits_{z_2}^{\infty} dz \rho(\mathbf{b}_1,z)\right\}~,
                                                                                 \label{PsurvD^*}
\end{eqnarray}
where $\rho=\rho_p+\rho_n$ is the total nucleon density. We use the two-parameter Fermi distributions of 
protons and neutrons \cite{Larionov:2013axa}.
As usual in the Glauber theory, Eqs.(\ref{Psurvpbar})-(\ref{PsurvD^*}) neglect the Fermi motion
of nucleon scatterers. In a similar way, in writing Eq.(\ref{dsigma_barpA_to_DsX}) we neglected the Fermi 
motion of nucleon 2 since the elementary cross section (\ref{dsigma_D^*_cov}) depends only weakly on the 
proton momentum (via the flux factors, screening- and antiscreening contributions) 
and is in leading order proportional to the square of the molecule wave function.
However, the transverse Fermi motion of proton 1 is taken into account 
in Eq.(\ref{dsigma_barpA_to_DsX}) in the high-energy approximation (c.f. \cite{Frankfurt:1979sv}).
The latter implies that the light cone momentum fraction $\alpha$ can be expressed in the target nucleus 
rest frame according to Eq.(\ref{alpha_Acm}) where the Fermi motion of the proton 1 is still neglected. 
(We have numerically checked that using the exact 
Lorentz transformation to the c.m. frame of $X$ to evaluate the invariant cross
section $\omega_{D^*}\frac{d^3 \sigma_{p X \to D^*}}{d^3 k}$ instead of using the infinite momentum 
frame in Eq.(\ref{dsigma_barpA_to_DsX}) which conserves $\alpha$ and assumes Galilean transformation 
for $\mathbf{k}_t$ produces indistinguishable results.)

The $\bar p p \to X$ matrix element in Eq.(\ref{Gamma_barp_to_X}) is one of the major uncertainties 
in our calculations. 
Its modulus squared can be formally expressed in terms of the partial decay width 
$\Gamma_{X \to \bar p p}$ as 
\begin{equation}
   \overline{|M_{X;\bar p 1}|^2}
   = \frac{4 \pi (2J_X+1) m_X^2 \Gamma_{X \to \bar p p}}{\sqrt{m_X^2-4m_N^2}}~, \label{MformAver}
\end{equation}
where the overline means averaging over antiproton and proton helicities and summation over the helicity
of $X$. There is no experimental data on the partial decay width $\Gamma_{X(3872) \to \bar p p}$.
(The recent LHCb data on $\bar p p$ invariant mass spectra 
from $B^+ \to p \bar p K^+$ decays \cite{Aaij:2013rha} do not allow to clearly
identify $X(3872)$ in the $p \bar p$ decay channel due to statistical limitations.)
In the present calculations, we will use the value $\Gamma_{X(3872) \to \bar p p} \simeq 30$ eV
as suggested by theoretical estimates \cite{Braaten:2007sh}.
This value is about two times smaller than $\Gamma_{\chi_{c1}(1P) \to \bar p p}$.
However, one should note that, in the molecular picture, the decay of the $X(3872)$ to the $p \bar p$ state
requires the production of only two $q \bar q$ pairs, and not three $q \bar q$ pairs as in the ordinary charmonium
decay to the $p \bar p$ channel. Thus, the partial decay width of the $X(3872)$ into the $p \bar p$
channel may be even larger than that of the $\chi_{c1}(1P)$ state \cite{Braaten:2007sh}.  

Formula (\ref{dsigma_barpA_to_DsX}) has a simple physical interpretation if we express the integral
$\int d^3 r_1$ as $\int d^2 b_1 \int dz_1$. The factor 
${\cal P}_{\bar p,{\rm surv}}(\mathbf{b}_1,-\infty,z_1)$ is the probability that the incoming
from $z=-\infty$ antiproton with impact parameter $\mathbf{b}_1$ will reach the point $z=z_1$.
The combination $(dz_1/v_{\bar p})d^2 p_{1t} d^2\Gamma_{\bar p}^{1 \to X}/d^2p_{1t}$
is the $X(3872)$ formation probability within the transverse momentum element
$d^2 p_{1t}$ when the $\bar p$ is passing the longitudinal element $dz_1$.
The factor ${\cal P}_{X,{\rm surv}}(\mathbf{b}_1,z_1,z_2)$ is the probability that the molecule
will reach the point $z=z_2$. The combination 
$dz_2 (d\alpha/\alpha) d^2k_t G_X^{p \to D^*}(\alpha,\mathbf{k}_t-\frac{\alpha}{2}\mathbf{p}_{1t})  
\rho(\mathbf{b}_1,z_2)$ is the probability that a $D^*$ will be produced in the kinematical element
$d\alpha d^2k_t$ when the $X(3872)$ is passing the longitudinal element $dz_2$.
Finally, the factor ${\cal P}_{D^*,{\rm surv}}(\mathbf{b}_1,z_2,\infty)$ is the probability 
that the $D^*$ will escape from the nucleus. In the spirit of the eikonal approach, all particles
propagate parallel to the beam direction. (For example, we assumed that the transverse momentum
of the molecule, $\mathbf{p}_{1t}$, does not influence its trajectory.)
The integration over $z_2$ can be taken with the explicit forms of the survival probabilities
Eqs.(\ref{PsurvX}),(\ref{PsurvD^*}). As a result Eq.(\ref{dsigma_barpA_to_DsX}) takes 
the following simple form:
\begin{eqnarray}
         && \alpha \frac{d^3\sigma_{\bar p A \to D^*}}{d\alpha d^2k_t} =
       \frac{1}{v_{\bar p}(\sigma_{pX}^{\rm tot}-\sigma_{pD^*}^{\rm tot})}
       \int d^3 r_1 {\cal P}_{\bar p,{\rm surv}}(\mathbf{b}_1,-\infty,z_1)
                 \int d^2 p_{1t} \frac{d^2\Gamma_{\bar p}^{1 \to X}}{d^2p_{1t}}
                 \nonumber \\  
   && \hspace{2cm} \times  G_X^{p \to D^*}(\alpha,\mathbf{k}_t-\frac{\alpha}{2}\mathbf{p}_{1t})
              [ {\cal P}_{D^*,{\rm surv}}(\mathbf{b}_1,z_1,\infty)
               -{\cal P}_{X,{\rm surv}}(\mathbf{b}_1,z_1,\infty) ]~. \label{dsigma_final}
\end{eqnarray}

\begin{figure}
\includegraphics[scale = 0.7]{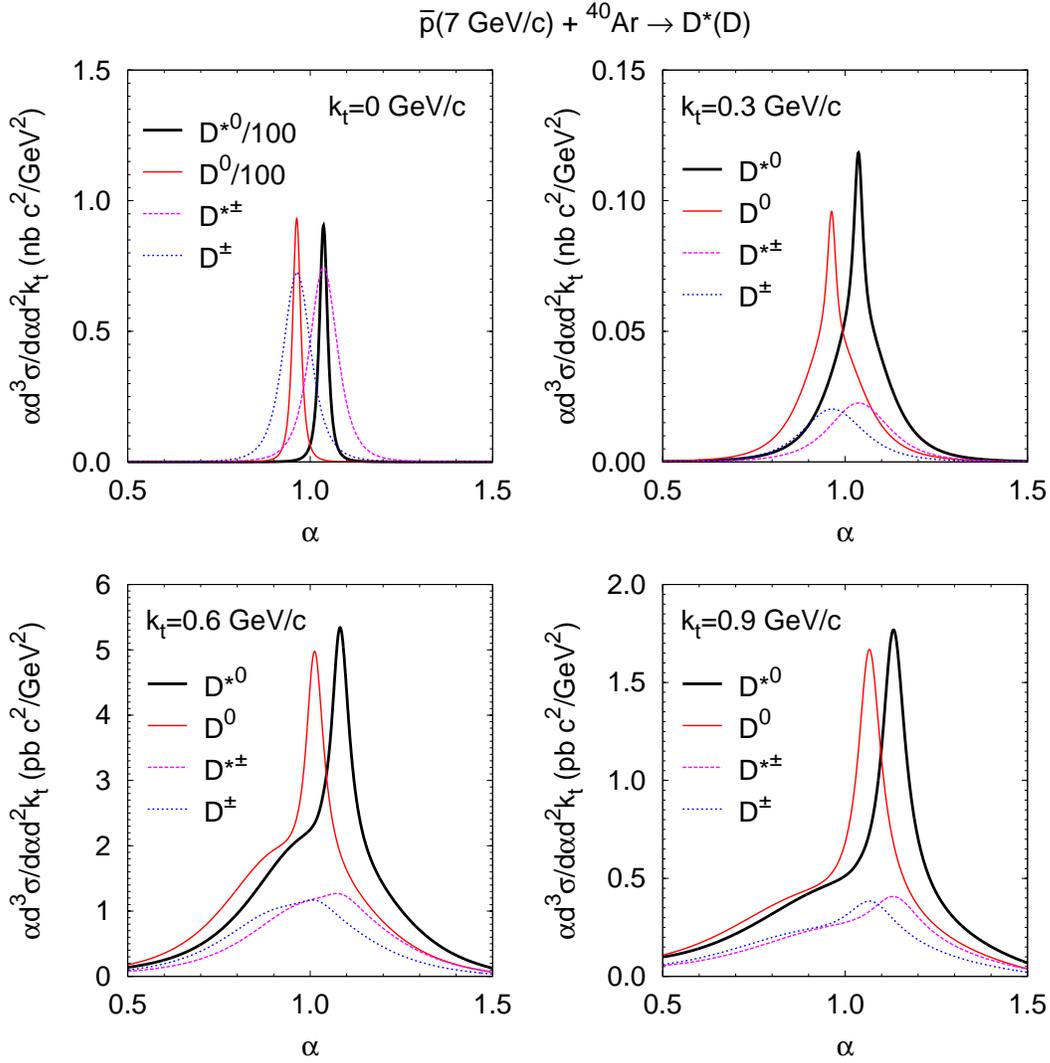}
\caption{\label{fig:pbarA_to_Ds} The invariant differential
cross sections of $D^{*0}$, $D^0$, $D^{*\pm}$ and $D^\pm$ production in $\bar p ^{40}$Ar collisions 
at $p_{\rm lab}=7$ GeV/c. The calculations are done using full cross section $X(3872) p \to D^*(D)$
in Eq.(\ref{dsigma_final}) including the IA term as well as screening and antiscreening 
corrections (see Eqs.(\ref{dsigma_D^*})-(\ref{dsigma_D^*_cov})). For $k_t=0$, the cross sections of 
$D^{*0}$ and $D^0$ production are divided by a factor of 100.}
\end{figure}
\begin{figure}
\begin{center}
\includegraphics[scale = 0.6]{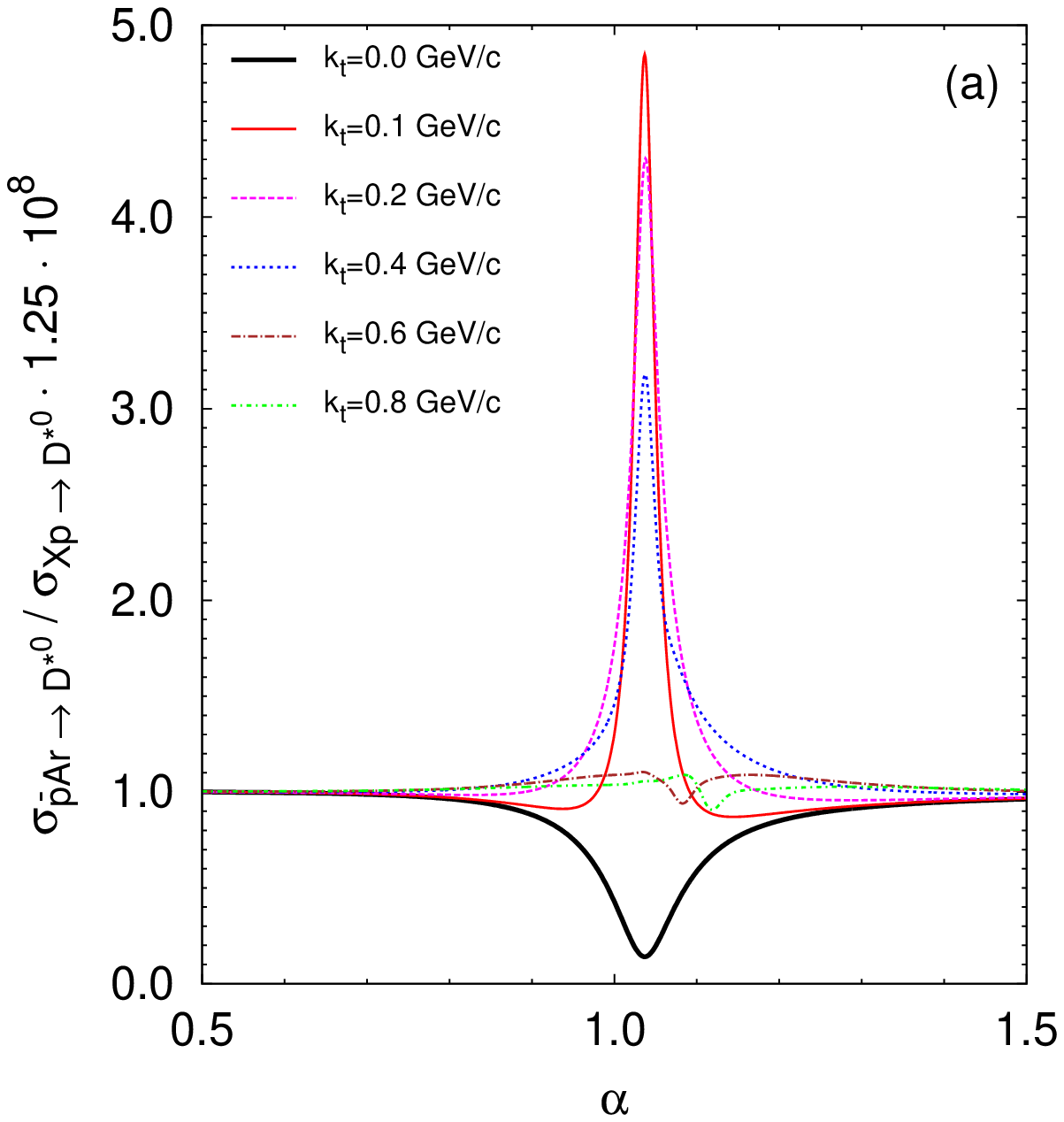}
\includegraphics[scale = 0.6]{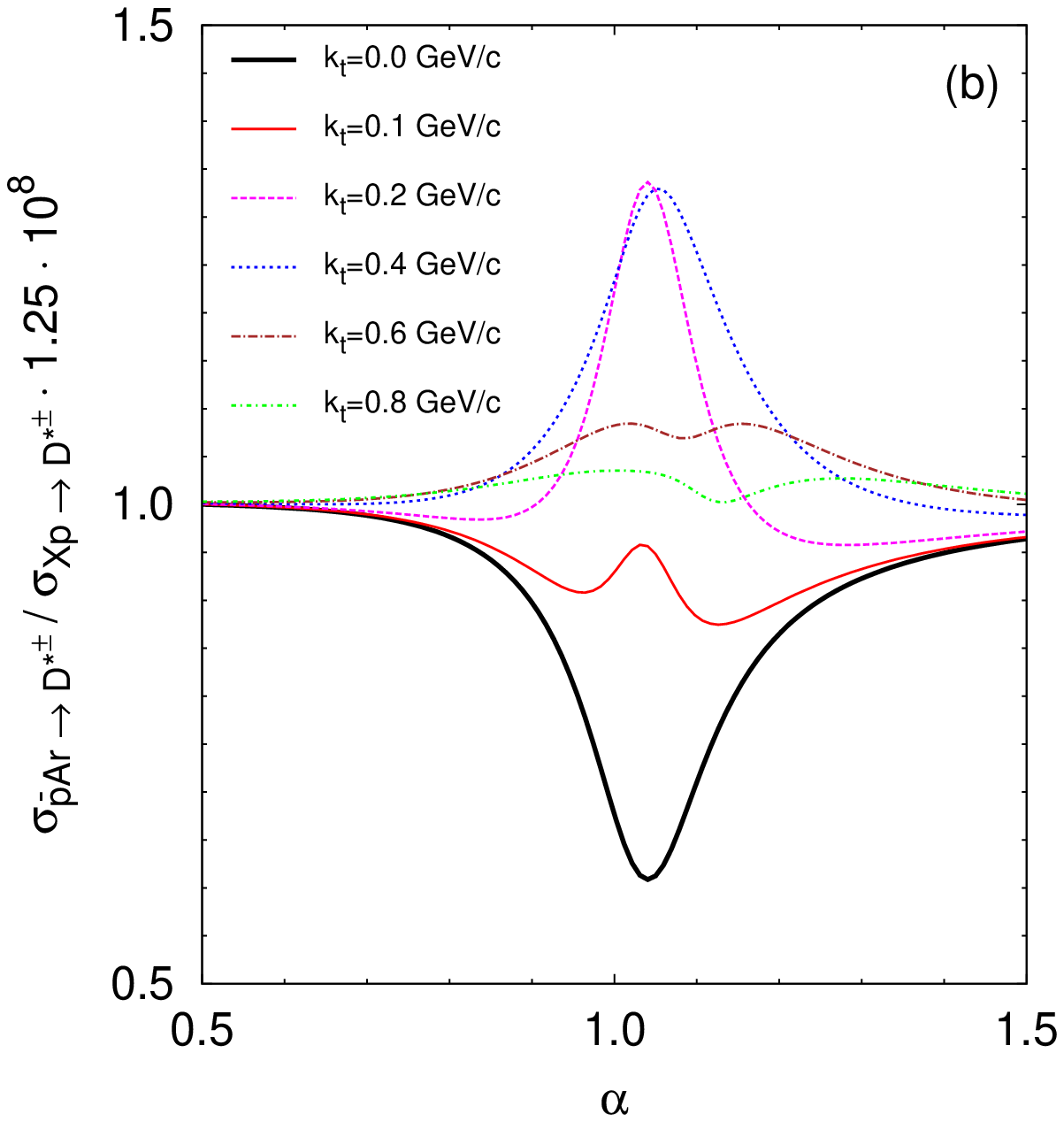}
\end{center}
\caption{\label{fig:ratio} The ratio of $D^{*}$ production cross sections
in $\bar p^{40}$Ar and $X(3872)p$ collisions at $p_{\rm lab}=7$ GeV/c 
for several values of transverse momentum $k_t$ as a function of light cone momentum fraction $\alpha$.
The ratio is normalized at unity for $\alpha=0.5$. Panel (a) -- $D^{*0}$. Panel (b) -- $D^{*\pm}$.}
\end{figure}

In Fig.~\ref{fig:pbarA_to_Ds} we display the differential cross sections of charmed meson
production in antiproton collisions with argon nucleus at 7 GeV/c.
The $D$ and $D^*$ cross sections are peaked at $\alpha \simeq 2m_D/m_X =0.96$ and
$\alpha \simeq 2m_{D^*}/m_X =1.04$, respectively, and behave in similar way
as a function of $\alpha$ and $k_t$. The widths of $\alpha$-dependence of the
$D^{*0}$ and $D^0$ cross sections are much smaller and the peak values are much larger as compared to the
$D^{*\pm}$ and $D^\pm$ cross sections. The $\alpha$-dependence of $D^*$ and $D$ production 
in $\bar p A$ collisions is dominated by the elementary cross section (c.f. 
Figs.~\ref{fig:Xp_to_Ds0},\ref{fig:Xp_to_Dsm}). However, a closer look reveals significant
differences between $D^*$ production on a nucleus and on a proton due to the Fermi motion.
These are better visible in the ratio of the two cross sections depicted in Fig.~\ref{fig:ratio}.
At $k_t=0$ the ratio has a minimum at $\alpha \simeq 2m_{D^*}/m_X$ because in this case
the contribution from target protons with finite transverse momentum $p_{1t}$ is
suppressed by the factor 
$|\psi(\mathbf{k}_t-\frac{\alpha}{2}\mathbf{p}_{1t})|^2/|\psi(\mathbf{k}_t)|^2$.
However, with increasing $k_t$ this factor becomes larger than unity for comoving proton 1.
This leads to the observed local maximum in the $\alpha$-dependence for $k_t \simeq 0.1-0.4$ GeV/c.
At large $k_t$ or for large deviations of $\alpha$ from unity the ratio tends to the constant value.

\section{Uncertainty and background}
\label{uncert}

\begin{figure}
\begin{center}
\includegraphics[scale = 0.4]{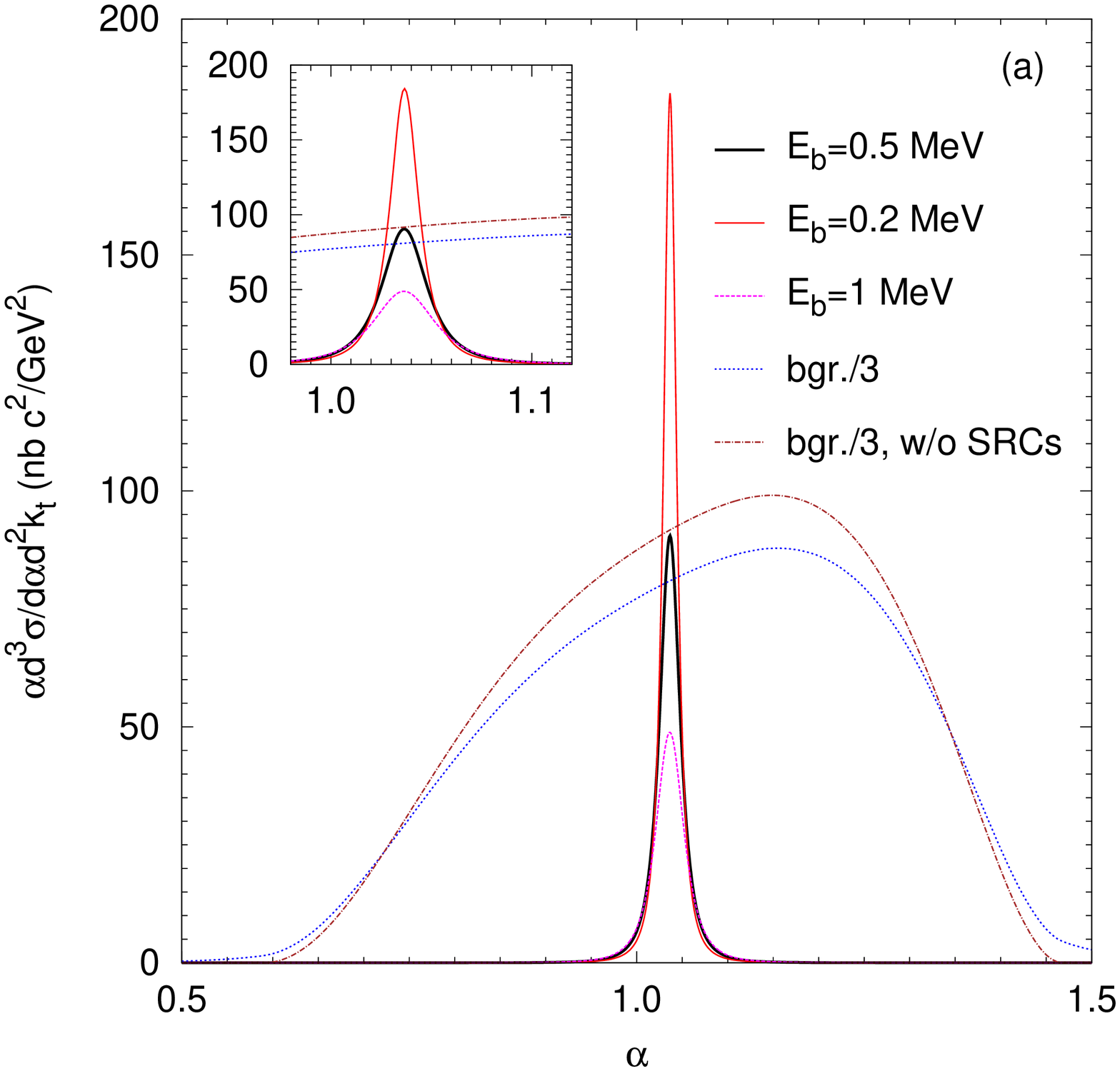}
\includegraphics[scale = 0.4]{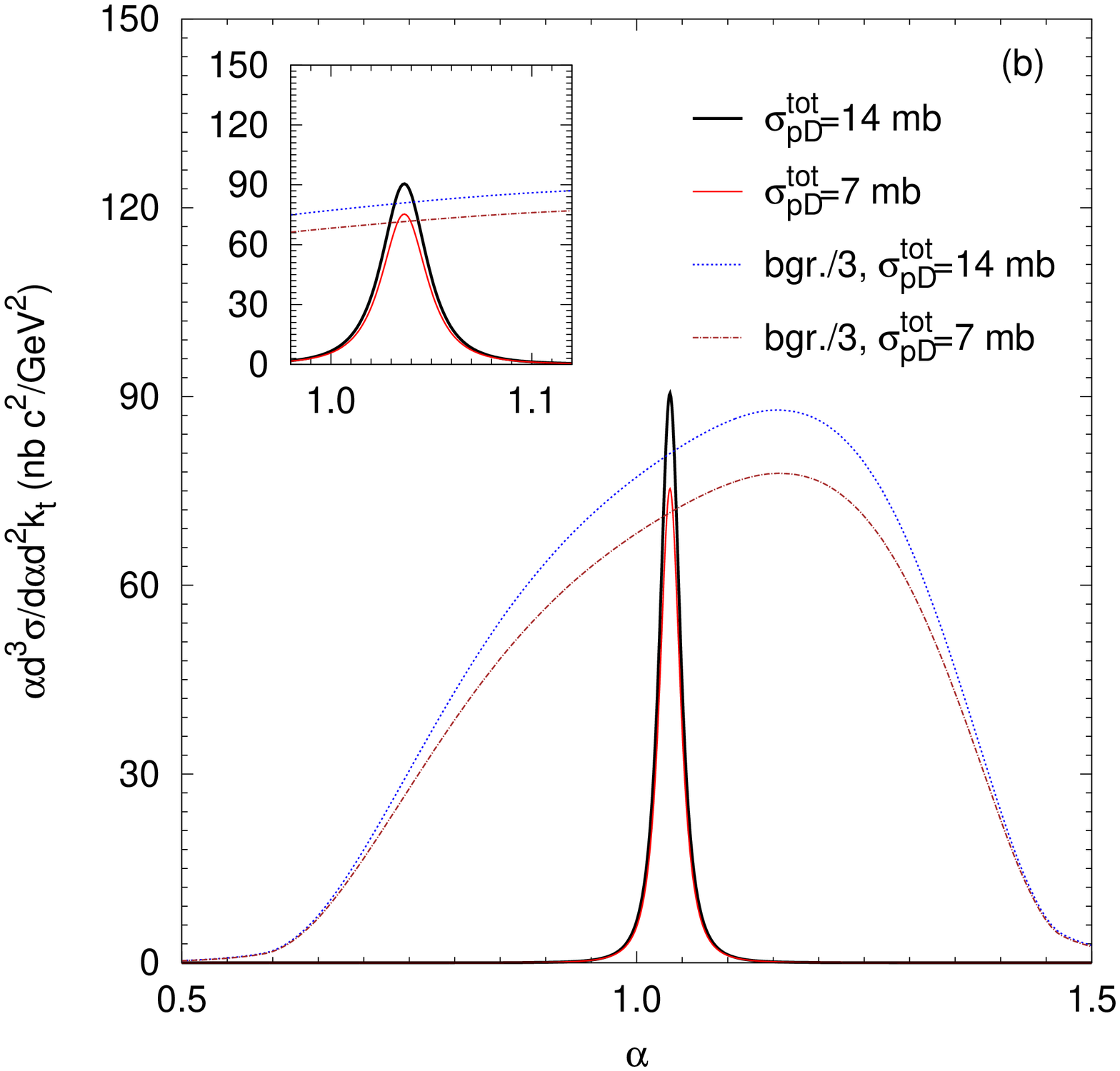}
\end{center}
\caption{\label{fig:err} The $\alpha$-dependence of $D^{*0}$ production at $k_t=0$ 
in $\bar p ^{40}$Ar collisions at $p_{\rm lab}=7$ GeV/c. Panel (a) shows calculations 
with default cross section  $\sigma_{pD}^{\rm tot}=14$ mb. The signal cross section (\ref{dsigma_final}) 
is shown for the different binding energies $E_b$ of the $D^0\bar D^{*0}$ molecule. The background
cross section (\ref{sig_bg}) is shown with SRCs (default calculation) and without SRCs.  
Panel (b) shows calculations with default $E_b=0.5$ MeV for the different $\sigma_{pD}^{\rm tot}$
as indicated. It is always assumed that $\sigma_{pD^*}^{\rm tot}=\sigma_{pD}^{\rm tot}$.
Insets show the narrower region of $\alpha$. The background cross section is divided by a factor of 3.}
\end{figure}
The uncertainty of our calculations can be read from Fig.~\ref{fig:err}. 
The unknown binding energy of the $D^0 \bar D^{*0}$ molecule is the main source of uncertainty 
as a weaker binding produces a narrower $\alpha$-distribution and vice versa.
As a consequence of the partial cancellation between the survival probabilities of $D^*$ and molecule 
the reduction of the $pD$ and $pD^*$ cross sections from 14 mb to 7 mb leads to the reduction of the peak 
of $D^{*0}$ production cross section by $\sim 15\%$ only. 
Of course, on the top of these two effects there is an uncertainty due to the experimentally unknown width       
$\Gamma_{X \to \bar p p}$ which enters the cross section (\ref{dsigma_final}) as an overall multiplication 
factor. 
The qualitative behavior of the $\alpha$-distribution is not changed by varying the
model parameters.

The major background is given by the direct process 
$\bar p N \to D \bar D^*, \bar D D^* $ on the bound nucleon, because the thresholds of $X(3872)$ and 
$D\bar D^*$ production in $\bar p p$ collisions are almost the same. 
The background cross section can be calculated as (derivation is similar to that of Eq. (\ref{dsigma_final})) 
\begin{eqnarray}
        && \alpha \frac{d^3\sigma^{\rm bg}_{\bar p A \to D^*}}{d\alpha d^2k_t} = \sum_{N_1=n,p}
           \frac{2}{(2\pi)^3 E_1 p_{\rm lab} k_D^z} 
           \int d^2 p_{1t}  \frac{q_{\bar p 1}(\varepsilon) \varepsilon^2}{q_{DD^*}(\varepsilon)} 
                           \frac{d \sigma_{\bar p 1 \to D^* D}(\varepsilon)}{d\Omega}       \nonumber \\
        && \hspace{3cm} \times \int d^3 r_1  {\cal P}_{\bar p,{\rm surv}}(\mathbf{b}_1,-\infty,z_1)     
                                             n_1(\mathbf{r}_1;\mathbf{p}_{1t},\Delta_\varepsilon^0)  
                                                  \nonumber \\ 
        && \hspace{4cm} \times {\cal P}_{D^*,{\rm surv}}(\mathbf{b}_1,z_1,\infty)
                        [1 - {\cal P}_{D,{\rm surv}}(\mathbf{b}_1,z_1,\infty)]~,     \label{sig_bg}
\end{eqnarray}
where
\begin{equation}
     \varepsilon \equiv \varepsilon(\alpha,\mathbf{k}_t^\prime =\mathbf{k}_t-\frac{\alpha}{2}\mathbf{p}_{1t})
    = \left(\frac{ 2[(2-\alpha)m_{D^*}^2 + \alpha m_D^2 + 2 k_t^{\prime 2}] }{ \alpha (2-\alpha) }\right)^{1/2}       \label{varepsilon}
\end{equation}
is the c.m. energy of $D$ and $D^*$; 
$q_{\bar p 1}(\varepsilon) = (\varepsilon^2/4-m_p^2)^{1/2}$ and
$q_{DD^*}(\varepsilon) = [( \varepsilon^2 + m_D^2 - m_{D^*}^2 )^2/4\varepsilon^2 -m_D^2]^{1/2}$
are the c.m. momenta of the colliding $\bar p N$ pair and of the produced $DD^*$ pair, respectively.
The longitudinal momentum $k_D^z$ of $D$-meson in the nucleus rest frame is calculated from relation
\begin{equation}
   2-\alpha=\frac{2(\omega_D+k_D^z)}{E_{\bar p}+m_N+p_{\rm lab}}~, \label{2malpha_Acm}
\end{equation}
with $\omega_D=\sqrt{{k_D^z}^2+m_D^2}$. The longitudinal momentum of the nucleon, $\Delta_\varepsilon^0$,
is given by Eq.(\ref{DeltaX^0}) with $m_X$ replaced by $\varepsilon$.

For the $D^{*0}$ production on the proton, the near-threshold $S$-wave cross section is
$\sigma_{\bar p p \to D^* D}(\varepsilon) = 2 \sigma_{\bar p p \to D^{*0} \bar D^0}(\varepsilon)$,
where the direct (non-resonant) cross section $\sigma_{\bar p p \to D^{*0} \bar D^0}(\varepsilon)$
has been taken from ref. \cite{Braaten:2007sh} which is the only estimate of the discussed 
cross section available in the literature. (We included the factor of 2 since our $D^{*0}$ 
includes both physical states, $D^{*0}$ and $\bar D^{*0}$.) 
The estimate of \cite{Braaten:2007sh} was obtained by using dimensional counting considerations 
to express the cross section of $\bar p p \to D^{*0} \bar D^0$ at high energies in terms of the cross section 
of $\bar p p \to K^{*-} K^+$ which is known in the limited energy range. As the next step, 
the $\bar p p \to D^{*0} \bar D^0$ cross section was extrapolated in \cite{Braaten:2007sh} towards the threshold 
and multiplied by the $S$-wave fraction $f_{L=0} \simeq 9.3\%$ which is regarded by the author of \cite{Braaten:2007sh} 
himself as ``a crude extrapolation''. Thus, we feel that the near-threshold estimate of \cite{Braaten:2007sh} 
can be considered as an order of magnitude estimate. 
In the case of the $D^{*0}$ production on the neutron, the only possible channel is $\bar p n \to D^{*0} D^-$. 
Thus, we assume $\sigma_{\bar p n \to D^* D}(\varepsilon) = \sigma_{\bar p p \to D^{*0} \bar D^0}(\varepsilon)$. 

The result of calculation using Eq.(\ref{sig_bg})
is shown in Fig.~\ref{fig:err}. The dependence of the background cross section on the
$pD^*$ and $pD$ cross sections is quite modest and follows the tendency of the signal cross 
section. Thus, at $k_t=0$, the sharp peak of $D^*$ production at $\alpha=1.04$ due to the stripping reaction
is clearly visible on the smooth background. The peak is almost not influenced by intramolecular 
screening and antiscreening effects. Moreover, we expect that the elastic rescattering of antiproton 
and produced particles on the nucleons will practically not change the $\bar D^*$ and $D$ spectra 
at small $k_t$ \cite{Larionov:2013nga}. Finally, the influence of the SRCs -- which are always included in default
calculations -- is mainly in the reduction of the nucleon occupancies at small momenta. This results in $15\%$ reduction 
of the background at $\alpha\simeq 1$ (Fig.~\ref{fig:err}a) and the correspoding rescaling of the signal cross section 
(not shown).

The signal-to-background ratio at the peak stays almost constant as the mass number of the target nucleus
varies between 20 and 208. The shape of the $\alpha$-dependence of the signal cross section practically 
does not vary with mass number in that region. The mass dependence of the total $D^{*0}$ production cross 
section due to the stripping reaction can be well approximated by formula $\sigma_{D^{*0}} =16~\mbox{pb} \cdot A^{0.46}$.
With the high luminosity mode at PANDA, $L=2 \cdot 10^{32}$ cm$^{-2}$ s$^{-1}$,  the estimated production 
rate due to the stripping reaction is about 60 $D^{*0}$ events per hour.

\section{Discussion and conclusions}
\label{discuss}

We proposed the idea of $D(\bar D^*)$ stripping from the $X(3872)$ state, to investigate if $X(3872)$
has a molecular structure from the narrow peaks in $\alpha$-distribution of $\bar D^*$ and $D$ at $\alpha \simeq 1$.
Other microscopic models of $X(3872)$, e.g. tetraquark or $c \bar c$-gluon hybrid, would
lead to the flat $\alpha$-spectrum of $\bar D^*$($D$). In such models, there are no primordial hadronic
components in $X(3872)$ to be ``released'' or ``knocked-out''. Thus, the momentum distribution 
of $\bar D^*$($D$) in the process $X N \to \bar D^*(D)$ would be dominated by phase space of the
final state particles.  

There is, however, another possible source of narrow peaks in $\alpha$-distributions of $\bar D^*$ and $D$.
The BELLE collaboration \cite{Adachi:2008sua} has found a significant near-threshold enhancement in the
$D^{*0} \bar D^0$ invariant mass spectrum from $B \to D^{*0} \bar D^0 K$ decays. 
We note that this does not exclude the existence of the $D^{*0} \bar D^0$ bound state.
(One similar example is $\Lambda(1405)$ which lies about 30 MeV below $K^- p$ threshold and can be
treated as a $K^- p$ quasibound state although it strongly influences the 
$K^- p \to \Sigma^{\pm} \pi^{\mp}$ and $K^- p \to \Sigma^0 \pi^0$ cross sections at small beam momenta 
\cite{Oset:1997it,Oller:2000fj}.)
But it is also possible that $X(3872)$ is a resonance coupled to the $D^{*0} \bar D^0 + c.c.$ channel.
If such a resonance state is produced in peripheral $\bar p A$ collisions, it will decay far away
from the nucleus, since the width of $X(3872)$ is less than 1 MeV.    
The resulting $\alpha$-distributions of $D^{*0}$ and $\bar D^0$ will be also sharply peaked near $\alpha \simeq 1$
at small $k_t$. However, in this case {\it both} decay products can in principle be detected.
This gives a clear experimental signature for distinguishing such decay events. In contrast, the stripping events
would contain {\it only one} meson, $D^{*0}$ or $\bar D^0$, in the same kinematical region.

The stripping reaction can be also considered for other production channels of $X(3872)$, e.g. in proton-,
electron- and photon-induced reactions on nuclei. 
Other exotic X,Y,Z states, such as the $X(3940)$ \cite{Abe:2007jn},
$Y(4140)$ \cite{Aaltonen:2009tz}, $X(4160)$ \cite{Abe:2007sya}
(c.f. recent reviews \cite{Godfrey:2008nc,Brambilla:2010cs} for a more complete 
list), may be interpreted as molecular states of $D^* \bar D^*$ or $D_s^* \bar D_s^*$.
These hypothetical molecular structures may also be tested by using stripping reactions, similar to $X(3872)$.
Experimentally, such studies could be performed at J-PARC, FAIR, SPS@CERN, and EIC. 

Apart from $c \bar c$ exotic states, there are other mesons which possibly have
molecular structures.
The $a_0(980)$ and $f_0(980)$ states viewed as $K \bar K$ bound state can be produced 
in $\gamma(\pi) N \to f N$ ($f \equiv a_0, f_0$) reaction on the bound nucleon 
followed by the stripping process $f N \to \bar K (K) + \mbox{anything}$ on another nucleon 
of the nuclear target residue.      
The $D_{s0}^*(2317)$ viewed as $DK$ and  $D_{s1}(2460)$ viewed as $D^*K$ can be produced in
$\bar p p \to D_s^\pm  D_{s0}^{*\mp} (D_{s1}^\mp)$ reaction on the bound proton
followed by the stripping process $D_{s0}^{*\mp} (D_{s1}^\mp) N \to D(D^*) + \mbox{anything}$
or $D_{s0}^{*\mp} (D_{s1}^\mp) N \to K + \mbox{anything}$ on another nucleon. The antiproton-nucleus
interactions open another unique opportunity to produce fast antibaryons. 
In this way, e.g. the molecular $K^+ \bar p$ hypothesis for the $\bar\Lambda(1405)$ can be tested
by using the two-step processes $\bar p p \to \bar\Lambda(1405) \Lambda$, 
$\bar\Lambda(1405) N \to K^+  + \mbox{anything}$. Such kind of processes can be studied at PANDA
and possibly at J-PARC.
 
In conclusion, we have demonstrated that the spectra of $\bar D^*$ and $D$ in the light cone momentum 
fraction at small transverse momenta allow to test the hypothetical $D\bar D^*$ molecular structure
of the $X(3872)$ produced in $\bar pA$ collisions at threshold. We propose to search the narrow peak
in $\bar D^*$ or $D$ production at $\alpha \simeq 1$ and small $k_t$ as an unambiguous signal of
the $D \bar D^*$ molecular state formation in $\bar p A$ collisions in PANDA experiment at FAIR.  

\section*{Acknowledgements}
\label{Ack}

M.S.'s research was supported by the US Department of Energy Office of Science, 
Office of Nuclear Physics under Award No.  DE-FG02-93ER40771. 
This work was supported by HIC for FAIR within the framework of the LOEWE program.

\bibliographystyle{apsrev}
\bibliography{pbarX3872}

\end{document}